\documentclass[a4paper,11pt]{article}

\usepackage{ragged2e} 
\usepackage{changepage}

\usepackage{siunitx} % degree symbol
\usepackage{amsmath}
\usepackage{bm} % bold in math environment
\usepackage{amssymb}
\usepackage{wasysym}
\usepackage{xfrac} % for other fraction (e.g. sfrac)

\usepackage{graphicx}
\usepackage{sidecap}
\usepackage{wrapfig}
\usepackage{caption}
\usepackage{subcaption}

\usepackage{verbatim}
\usepackage[english]{babel}

\usepackage{array}
\usepackage{adjustbox}
\usepackage{float}
\usepackage{enumerate}
\usepackage{multirow}
\usepackage{multicol}
\usepackage[table,xcdraw]{xcolor}
\usepackage{natbib}

\usepackage{csvsimple}
\usepackage{bbm}
\usepackage{pdflscape}
\usepackage{breqn}
\usepackage{lscape}
\usepackage{amsfonts}
\usepackage{latexsym}
 \usepackage[titletoc]{appendix}  % for splitting a table of contents in 2 pages
\usepackage[colorlinks=true, citecolor= black]{hyperref} % hyperlinks
\hypersetup{linkcolor=black} % hyperlinks
\usepackage[all]{hypcap} % hyperlinks
\usepackage{setspace}
\usepackage{enumitem}
\usepackage{authblk}
\usepackage{bm}

\title{\textbf{Federated mixed effects logistic regression based on one-time shared summary statistics}}
\author[1]{\small Marie Analiz April Limpoco}
\author[1]{\small Christel Faes}
\author[1,2]{\small Niel Hens}
\affil[1]{Interuniversity Institute for Biostatistics and Statistical Bioinformatics (I-BioStat), Data Science Institute (DSI), Hasselt University, Hasselt, Belgium}
\affil[2]{Centre for Health Economic Research and Modelling Infectious Diseases (CHERMID), Vaccine \& Infectious Disease Institute, University of Antwerp, Antwerp, Belgium}
\date{}

\newtheorem{theorem}{Theorem}[section]
\begin{document}
\maketitle

\begin{abstract}
    Upholding data privacy especially in medical research has become tantamount to facing difficulties in accessing individual-level patient data. Estimating mixed effects binary logistic regression models involving data from multiple data providers like hospitals thus becomes more challenging. Federated learning has emerged as an option to preserve the privacy of individual observations while still estimating a global model that can be interpreted on the individual level, but it usually involves iterative communication between the data providers and the data analyst. In this paper, we present a strategy to estimate a mixed effects binary logistic regression model that requires data providers to share summary statistics only once. It involves generating pseudo-data whose summary statistics match those of the actual data and using these into the model estimation process instead of the actual unavailable data. Our strategy is able to include multiple predictors which can be a combination of continuous and categorical variables. Through simulation, we show that our approach estimates the true model at least as good as the one which requires the pooled individual observations. An illustrative example using real data is provided. Unlike typical federated learning algorithms, our approach eliminates infrastructure requirements and security issues while being communication efficient and while accounting for heterogeneity. \\

\end{abstract}

\textbf{Correspondence:} Marie Analiz April Limpoco (liz.limpoco@uhasselt.be)

\textbf{Funding:} This study was supported by the Special Research Fund of Hasselt University (BOF24OWB22, Methusalem grant).

\doublespacing
%-----------INTRODUCTION-----------
\section{Introduction}
Binary logistic regression has widespread use in health research and beyond \citep{BAGLEY2001979, https://doi.org/10.1111/jpc.12895, Harrise001290, ZABOR2022271}. It allows for the assessment of risks, such as for example the association between the presence of a disease and one or more factors (e.g. age, sex, socioeconomic factors). Logistic mixed models are used when the data are clustered or have a hierarchical structure. For example, when data come from multiple hospitals, adding a random effect will account for possible heterogeneity among hospitals, making the inference and prediction more reliable \citep{https://doi.org/10.1002/bimj.201900075}. 

However like any statistical model, logistic regression models and logistic mixed models require individual-level data in the model estimation process. With the increasing emphasis on data privacy, analysing individual level data through statistical modelling becomes more challenging both for the data providers and data analysts \citep{https://doi.org/10.1111/j.1475-6773.2010.01141.x}. Indeed, for example, implementing logistic mixed models across multiple hospitals poses challenges, as hospital data are often sensitive, containing patient information protected by strict privacy regulations. This creates a tension between the need for rich fine-scale data for accurate modeling and the necessity of preserving patient privacy.\\

One possible solution is data aggregation, where each data provider computes and shares aggregated summary statistics instead of individual-level data.
For models with categorical predictors only, this can yield still the same estimates as those derived from individual-level data \citep{https://doi.org/10.1002/bimj.201900034, Li2021}. Specifically, the data can be collapsed into counts within unique covariate combinations and the log-likelihood can be rewritten in terms of this summary data only. However, this approach struggles when there are continuous covariates. \cite{https://doi.org/10.1002/bimj.201900034} suggest to perform coarsening (e.g. by categorizing the continuous variable) \citep{f478dc7b-0e89-31b2-9241-5938769077a4}, but this introduces bias \citep{marshall_2016} and leads to loss of information \citep{FELSENSTEIN199899}. In addition, results depend on the degrees of data aggregation \citep{GARRETT200361}. This difficulty is attributed to the non-existence of exact sufficient statistics for binary logistic regression models involving continuous predictors, unlike linear regression models \citep{Huggins2017PASSGLMPA}. For linear mixed models, \cite{limpoco2024linearmixedmodellingfederated} showed that it is sufficient to have just the mean, covariance, and sample size from each data provider to estimate exactly the same model as based on the individual-level data.  
\\

To overcome these limitations, federated learning has emerged as a promising solution for fitting models using data from multiple hospitals, without requiring the transfer of individual-level data. Federated learning algorithms allow each hospital to fit a local model using its own data while iteratively sharing only intermediate updates with a central server \citep{50116, liu2020systematic, 45648, BANABILAH2022103061}. These methods require infrastructure that ensures secured communication between the data providers and data analysts without accessing the individual observations, making its practical implementation challenging \citep{9084352}. For generalized linear mixed models (GLMM), including logistic regression with random intercepts, federated learning algorithms have already been developed \citep{10.1371/journal.pone.0280192, Li2022}. These algorithms rely on communication-efficient strategies, such as distributed versions of the penalized quasi-likelihood method, to minimize the need for extensive communication with the data providers \citep{10.1093/jamia/ocac067}. However, these methods still face challenges, including the computational burden of iteratively sharing updates and potential biases associated  with using the penalized quasi-likelihood approach \citep{3c3e7e1e-657d-3e0a-a61d-6012f81dcaeb, Li2022}.  \\

To further reduce the need for iterative communications with the data providers, recent work has explored non-iterative methods. For example, an approach that accommodates both categorical and continuous covariates is based on artificial data generation using a Gaussian copula \citep{https://doi.org/10.1002/sim.8470}. This strategy makes use of empirical marginal moments and correlation matrices as specification for the marginal and joint components of the Gaussian copula, from which several sets of artificial data are simulated. Parameters are estimated from each set and then aggregated to come up with a single set of parameter estimates. Similarly, surrogate likelihoods allow for model estimation using gradients of the likelihood function \citep{doi:10.1142/9789813279827_0004, 10.1093/jamia/ocz199}. However, it requires access to the full individual-level data of at least one data provider, or alternatively has the capacity to share gradients of the likelihood amongst the data providers. If this is not possible, a compromise is to allow iterative communication of intermediary results until the global model converges \citep{10.1136/amiajnl-2012-000862},  \citep{a15070243}. \\

The non-iterative method proposed in this paper builds on these ideas by generating pseudo-data that approximate the summary statistics of the original data, allowing for the estimation of logistic mixed models without requiring iterative communication. This method can accommodate multiple covariates which can be both continuous or categorical. It employs adaptive Gauss-Hermite quadrature (implemented in \texttt{R}) which is considered more accurate in approximating the likelihood of a GLMM compared to the penalized quasi-likelihood-based method \citep{book}. Through simulations, it is shown that the estimated federated model performs at least as good as the classical method based on the actual individual-level data. This method overcomes the limitations from coarsening continuous variables and iterative communication and provides an accurate and privacy-preserving solution for multi-hospital data analysis.\\

In the next section, the details of the method as well as the simulation experiments that assess the performance of the model estimation process are presented. Thereafter, the simulation results are shown, followed by a real-world example that demonstrates the proposed strategy using publicly available de-identified and anonymized data from the Children's Hospital of Pennsylvania (CHOP) \href{https://higgi13425.github.io/medicaldata/}{(https://higgi13425.github.io/medicaldata/)}. Lastly, the relevant implications, insights and final conclusions are discussed.

%-----------METHODS-----------
\section{Theory and methods}
The proposed strategy requires the data providers (e.g. hospitals) to supply the data analyst with privacy-preserving summary statistics. An underlying assumption is that the variables are defined and measured consistently across data providers. Additionally, these summary statistics are assumed to be computed from complete cases, i.e. there are no missing values. From these, the data analyst first generates ``pseudo-data'' and subsequently uses the pseudo-data to estimate a binary logistic regression or a mixed effects logistic regression model. In this section, we provide details about the required summary statistics (Section 2.1) and how these are sufficient for reliable model estimation. Moreover, we describe how to generate pseudo-data from these summary statistics through unconstrained nonlinear least squares optimization (Section 2.2). To assess the performance of the proposed approach, a simulation study is conducted (Section 2.5). \\

\subsection{Polynomial-approximate sufficient statistics}\label{polsuf}
 In the setting where access to individual-level data is restricted due to privacy reasons, supplying sufficient statistics to the data analyst allows for model estimation to proceed with just one round of communication. However, unlike linear regression models, binary logistic regression models do not have exact sufficient statistics. Despite this, we will show that certain summary statistics such as the mean and sample central moments act like sufficient statistics up to a certain level of accuracy. \\

Consider the log-likelihood function of a binary logistic regression without random effects 
\begin{align*}
            l(\boldsymbol{\beta}) &= \sum_{i=1}^n{y_i\mathbf{x}_i^T\boldsymbol{\beta}} - {{\sum_{i=1}^n{\text{log}(1 + \text{exp}(\mathbf{x}_i^T\boldsymbol{\beta}))}}},
\end{align*}
where $y_i$ is the binary response of individual $\{i\}_{1}^n $, $\mathbf{x}_i$ is the $(p+1)$-dimensional vector of predictor data for individual $i$ with $p$ predictors, and $\boldsymbol{\beta}$ denotes the $(p+1)$-dimensional vector of coefficients (including the intercept) to be estimated. The first term of the log-likelihood can utilize the aggregated form $\sum_{i=1}^n{y_i\mathbf{x}_i^T}$ and thus does not necessitate individual-level data $y_i$ and $\mathbf{x}_i$ to be disclosed. In contrast, the second term is difficult to disentangle because of the logarithmic and exponential functions within the summand.  Additionally, because the maximum likelihood estimator for this model does not have a closed form, coefficients are estimated in an iterative manner until convergence. Consequently, utilizing all individual observations for every iteration has been the classical means of estimation, which is challenging when dealing with sensitive data. To circumvent this, we apply the following theorem in mathematical analysis: 

\begin{theorem}[Weierstrass approximation theorem]
    Suppose $f$ is a continuous real-valued function defined on the real interval $[a,b]$. For every $\varepsilon > 0$, there exists a polynomial $p$ of degree $K$ such that for all $x$ in $[a,b]$, we have $|f(x)-p(x)| < \varepsilon$.
\end{theorem}

Let $f(\mathbf{x}_i^T\boldsymbol{\beta}) = {\text{log}(1 + \text{exp}(\mathbf{x}_i^T\boldsymbol{\beta}))}$ be the continuous function defined on some finite interval $[a,b]$. Performing algebraic manipulations results in the second term of the log-likelihood being expressed as

\begin{align}
    \sum_{i=1}^n f(\mathbf{x}_i^T\boldsymbol{\beta}) &= \sum_{i=1}^n \left(\sum_{k=0}^K c_k(\mathbf{x}_i^T\boldsymbol{\beta})^k + \varepsilon_i\right) \\
    &= \sum_{k=0}^K c_k \left(\sum_{i=1}^n (\mathbf{x}_i^T\boldsymbol{\beta})^k \right) + \sum_{i=1}^n \varepsilon_i,
\end{align}

where $c_k$ is the coefficient corresponding to each monomial term $(\mathbf{x}_i^T\boldsymbol{\beta})^k$ of the polynomial of degree $K$, while $\varepsilon_i > 0$ is the error of approximating $f$ using the polynomial. This form of the second term of the log-likelihood shows the possibility of identifying sufficient statistics up to a certain level of accuracy $\varepsilon = \sum_{i}\varepsilon_i > 0$ of the polynomial approximation. We use the term \textit{polynomial-approximate sufficient statistics} to refer to these summary statistics to distinguish them from exact sufficient statistics. \\

Among the different methods of polynomial approximation, Taylor polynomial approximation leads to the use of sample central moments as polynomial-approximate sufficient statistics. This is appealing intuitively because moments describe properties of probability distributions such as central tendency, spread, symmetry, and shape, to name a few. Taylor's theorem states that:

\begin{theorem}[Taylor's theorem]
    Suppose $f$ is defined on some open interval $I$ around a real number $\text{a}$ and is at least $K+1$-times differentiable on this interval. Then for each $x \neq a$ in~$I$, there exists a value $\xi$ between $x$ and $a$ such that
    \begin{align}
        f(x) &= \sum_{k=0}^K \frac{f^{(k)}(a)}{k!}\left(x-a\right)^k + \frac{f^{(K+1)}(\xi)}{(K+1)!}\left(x-a\right)^{K+1}.
    \end{align}
\end{theorem}

If we rewrite the second term of the log-likelihood as a function of $\mathbf{x}_i^T\boldsymbol{\beta}$ and choose the center to be $a = \mathbf{\bar{x}}^T\boldsymbol{\beta}$ where $\mathbf{\bar{x}}$ is the vector of sample means of the $p$ regressors:

\begin{align}\label{taylor}
    \sum_{i=1}^n f(\mathbf{x}_i^T\boldsymbol{\beta}) &= \sum_{i=1}^n \left(\sum_{k=0}^K \frac{f^{(k)}(\mathbf{\bar{x}}^T\boldsymbol{\beta})}{k!}\left(\mathbf{x}_i^T\boldsymbol{\beta}-\mathbf{\bar{x}}^T\boldsymbol{\beta}\right)^k + \frac{f^{(K+1)}(\xi_i)}{(K+1)!} \left(\mathbf{x}_i^T\boldsymbol{\beta}-\mathbf{\bar{x}}^T\boldsymbol{\beta}\right)^{K+1}\right) \\
    &= \sum_{k=0}^K \frac{f^{(k)}(\mathbf{\bar{x}}^T\boldsymbol{\beta})}{k!}\sum_{i=1}^n \left((\mathbf{x}_i - \mathbf{\bar{x}})^T\boldsymbol{\beta}\right)^k + \sum_{i=1}^n \frac{f^{(K+1)}(\xi_i)}{(K+1)!} \left((\mathbf{x}_i - \mathbf{\bar{x}})^T\boldsymbol{\beta}\right)^{K+1},
\end{align}

where $\xi_i$ is a value between  $\mathbf{x}_i^T\boldsymbol{\beta}$ and $\mathbf{\bar{x}}^T\boldsymbol{\beta}$. This form suggests using the sample central moments, both univariate and joint, aside from the mean, as polynomial-approximate sufficient statistics. \\

The univariate sample central moment of order $r$ of a variable $X$ is defined as 
\begin{align}\label{mu_r}
    \bar{\mu}^{(r)} &= \frac{1}{n}\sum_{i=1}^n{(x_i - \bar{x})^r}, \hspace{5mm} r = 1,2,...
\end{align}

For example, the $2$nd sample central moment is the variance
\begin{align}
    \bar{\mu}^{(2)} &= \frac{1}{n}\sum_{i=1}^n{(x_i - \bar{x})^2}.
\end{align}

The bivariate or joint sample central moment of order $(r_1,r_2)$ between variables $X_1$ and $X_2$ is
\begin{align}
    \bar{\mu}^{(r_1,r_2)} &= \frac{1}{n}\sum_{i=1}^n{(x_{1i} - \bar{x}_1)^{r_1}(x_{2i} - \bar{x}_2)^{r_2}}, \hspace{5mm} r_1, r_2 = 1,2,...
\end{align}

The sample covariance between two variables is the $(1,1)$-order joint sample central moment between $X_1$ and $X_2$:
\begin{align}
    \bar{\mu}^{(1,1)} &= \frac{1}{n}\sum_{i=1}^n{(x_{1i} - \bar{x}_1)(x_{2i} - \bar{x}_2)}.
\end{align}

In general, the $p$-variate sample central moment of order $\mathbf{r} = (r_1,r_2,...,r_p)$, using a multi-index notation, is
\begin{align}
    \bar{\mu}^{(\mathbf{r})} &= \frac{1}{n}\sum_{i=1}^n {(x_{i} - \bar{x})^{\mathbf{r}}},
\end{align}

where $|\mathbf{r}| = r_1 + r_2 + ... + r_p$ and $x^{\mathbf{r}} = x_1^{r_1} x_2^{r_2}...x_p^{r_p}$. For instance, the bivariate sample central moments of order $|\mathbf{r}| = 3$ are:
\begin{align}
    \bar{\mu}^{(1,2)} &= \frac{1}{n}\sum_{i=1}^n {(x_{1i} - \bar{x}_1)(x_{2i} - \bar{x}_2)^2}, \hspace{3mm} \text{and} \\
    \bar{\mu}^{(2,1)} &= \frac{1}{n}\sum_{i=1}^n {(x_{1i} - \bar{x}_1)^2(x_{2i} - \bar{x}_2)}.
\end{align}

These polynomial-approximate sufficient statistics are the summary statistics expected from each data provider to proceed with model estimation without requiring the disclosure of individual-level data. For a binary variable, we follow the same formulas for computing the central moments by virtue of the Taylor polynomial expansion shown in Equation \ref{taylor}. Meanwhile, categorical predictors with more than two levels are converted first into the appropriate dummy variables, thereby turning them into binary variables, before computing the central moments.

\subsection{Pseudo-data generation}\label{genpseudo}
\begin{figure}[h]
        \centering
        \includegraphics[width=0.75\linewidth]{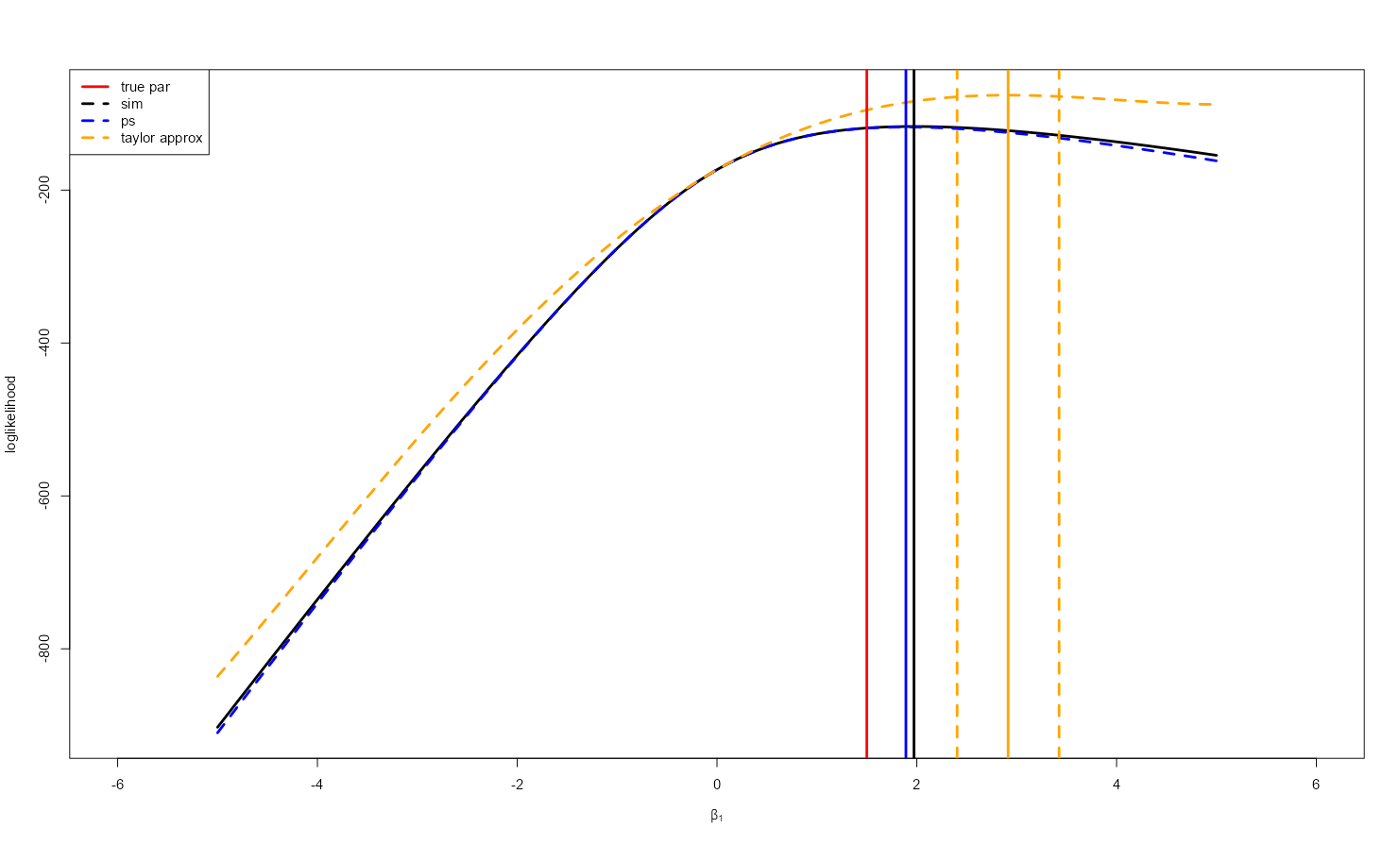}
        \caption{Comparison of log-likelihoods constructed from simulated data (sim) serving as the actual inaccessible data, pseudo-data (ps) generated from the polynomial-approximate sufficient statistics of the simulated data, and from polynomial approximation (taylor approx). The red solid vertical line (true par) represents the true parameter value of $\beta_1$ while the black (sim) and blue (ps) solid vertical lines represent the ML point estimates derived from the simulated and pseudo-data, respectively. The orange solid vertical line represents the ML point estimate derived from maximizing the polynomial-approximated log-likelihood (taylor approx) while the orange dashed vertical lines are the corresponding interval estimates \href{https://lizlimpoco.shinyapps.io/approx_loglik/}{(https://lizlimpoco.shinyapps.io/approx\_loglik/)}.} 
        \label{fig:approx_loglik}
    \end{figure}

We propose generating pseudo-data from the provided polynomial-approximate sufficient statistics and using them in place of the actual data instead of directly using the polynomial-approximate sufficient statistics in estimating the model. We define pseudo-data as artificially generated data with the same polynomial-approximate sufficient statistics as the actual data. This strategy is proposed because using a polynomial-approximated log-likelihood alone by assuming that the error part is zero may result in a function that does not closely resemble the actual log-likelihood. Moreover, the concavity and shape of this function might affect the maximization process, leading to estimates that are too far from the truth, or to numerical problems. Figure \ref{fig:approx_loglik} exemplifies this possibility, wherein the polynomial-approximated log-likelihood clearly deviates from the actual log-likelihood, along with the generated estimates. On the contrary, the log-likelihood constructed from the pseudo-data closely resembles the actual log-likelihood. It follows that the estimates are also very similar to the actual and are close to the true parameter value as well. \\

With the goal of using pseudo-data in place of the unavailable actual individual-level data, it is crucial to match salient statistical properties between them. We have shown that for a binary logistic regression, the log-likelihood can be rewritten in terms of the polynomial-approximate sufficient statistics. It follows that if the actual and pseudo-data have equal summary statistics in the form of  these polynomial-approximate sufficient statistics, the corresponding log-likelihoods will be similar. \\ 

To match the summary statistics, we apply the concept of unconstrained nonlinear least squares optimization, which does not impose any distributional assumptions. Our objective is to minimize the sum of squared differences (SSD) in the summary statistics between the pseudo-data and actual data. For a model with one predictor, let the sample size be equal to $n$, $\mathbf{y}_d$ be the vector of actual binary responses, and $\mathbf{x}_d$ be the vector of actual predictor values. Both $\mathbf{y}_d$ and $\mathbf{x}_d$ are not accessible, but their summary statistics namely the mean of both ($\bar{\mathbf{y}}_d$, $\bar{\mathbf{x}}_d$) and the variance ($s^2_{d}$), covariance ($s_{d_{xy}}$), and $r$th central moment ($\bar{\mu}^{(r)}_{d}$) of $\mathbf{x}_d$ are. To obtain the $n$-dimensional vector $\mathbf{x}_\pi$ of pseudo-data values, we minimize the following objective function: 
\begin{align*}
       SSD &= (\bar{\mathbf{x}}_{d} - \bar{\mathbf{x}}_{\pi})^2 + (s^2_{d} - s^2_{\pi})^2 + (s_{d_{xy}} - s_{\pi_{xy}})^2 + (\bar{\mu}^{(3)}_{d} - \bar{\mu}^{(3)}_{\pi})^2 + \hdots + (\bar{\mu}^{(r)}_{d} - \bar{\mu}^{(r)}_{\pi})^2. 
\end{align*}
Here, the summary statistics with subscript $d$ are supplied by the data provider while those with a subscript $\pi$ implicitly contain the unknown $\mathbf{x}_\pi$ pseudo-data values. Hence, optimization is done with respect to the $n$-dimensional vector $\mathbf{x}_\pi$. \\

In this paper, we consider sample central moments up to $r=4$ for practical reasons: they are easier to compute than higher-order moments and they have physical meaning concerning the location, spread, and shape of the distribution \citep{DANG2019505}. We implement the Levenberg-Marquardt method to solve this optimization problem. This method is well-known and prevalent in curve-fitting wherein the parameters of a model are estimated based on a set of data points \citep{83b09f23-b20e-3617-8f72-24765b713f7b, 225dd72e-23e2-30d0-ad93-6fc234f74f64, ortega2000iterative, dennis1996numerical, kelley1995iterative, KANZOW2004375}.  More often than not, the size of pseudo-data $n$ is larger than the number of supplied summary statistics, resulting in an underdetermined system. This implies that there are infinitely many solutions to the optimization problem, meaning that there are infinitely many possible sets of pseudo-observations that correspond to the same set of summary statistics. Consequently, it is improbable to exactly reconstruct the actual data, which is favorable in terms of upholding privacy.\\

Whenever there are two or more predictors, joint moments are included in the optimization problem. For instance, generating pseudo-data that matches up to the 3rd order central moments of the actual data with two predictors requires the bivariate sample central moments of order $(1,1)$ (covariance), $(1,2)$, and $(2,1)$. These are in addition to each predictor's univariate 3rd moment, variance, covariance with the response variable, mean, and sample size. Moreover, we generate the pseudo-observations one variable at a time to reduce the computational burden associated with optimizing several unknowns simultaneously. Note that for binary predictor variables, the generated pseudo-data may no longer be binary. \\

In summary, we propose the following steps to generate pseudo-data:
\begin{enumerate}
    \item Generate binary pseudo-data for the response variable ($\mathbf{y}_{\pi}$) based on the sample proportion of events $\mathbf{\bar{y}}_d$.
    \item For predictor $j = 1, 2, \hdots, p$, generate pseudo-data $\mathbf{x}_{\pi_j}$ based on the summary statistics of the actual data $\mathbf{x}_{d_j}$.
\end{enumerate}
 
 We propose to generate the pseudo-responses first and ensure that they are strictly binary. This enables the use of the existing statistical functionalities for logistic regression model estimation once the pseudo-data are generated. These functionalities constrain the response to be binary while the predictor type is unconstrained. The code to generate pseudo-data was developed in \texttt{R} \citep{Rpackage} via the \texttt{lsqnonlin} function from the package \texttt{pracma} \citep{Rpracma} and is available in our Github repository \href{https://github.com/lizlimpocouhasselt/Mixed-effects-logistic-regression-from-summary-statistics}{(https://github.com/lizlimpocouhasselt/Mixed-effects-logistic-regression-from-summary-statistics)}. 

\subsection{Mixed effects logistic regression}\label{mixedglm}
It can be shown that the concepts of polynomial-approximate sufficient statistics and pseudo-data generation are applicable whenever random effects are present in the model. We limit the complexity of our model in this paper to two levels but the logic is the same for more levels. \\

Let $y_{ij}$ denote the binary response of the $j$th individual ($j=1,2,\hdots,n_i$) from cluster $i$ ($i=1,2,\hdots,m$) while $\mathbf{x}_{ij}$ is the $(p+1)$-dimensional vector of predictor data corresponding to the vector of fixed effects $\boldsymbol{\beta}$ whose first element is the intercept. Further denote the $q$-dimensional vector of random effects associated with cluster $i$ to be $\mathbf{u}_i$ while $\mathbf{z}_{ij}$ is the corresponding vector of data. The conditional probability of the response vector for cluster $i$ represented by $\mathbf{y}_i=[y_{i1},y_{i2},\hdots,y_{in_i}]^T$ is
\begin{align}
    g(\mathbf{y}_i|\mathbf{u}_i;\boldsymbol{\beta}) &= \prod_{j=1}^{n_i} \left[p(y_{ij}=1|\mathbf{u}_i)\right]^{y_{ij}}\left[1-p(y_{ij}=1|\mathbf{u}_i)\right]^{1-y_{ij}},
\end{align}
where $p(y_{ij}=1|\mathbf{u}_i)=\frac{\exp{(\mathbf{x}_{ij}^T\boldsymbol{\beta} + \mathbf{z}_{ij}^T\mathbf{u}_i)}}{1+\exp{(\mathbf{x}_{ij}^T\boldsymbol{\beta} + \mathbf{z}_{ij}^T\mathbf{u}_i)}}$. It can be shown that 
\begin{align}
    g(\mathbf{y}_i|\mathbf{u}_i;\boldsymbol{\beta}) &= \exp{\left(\log{\left(\prod_{j=1}^{n_i} \left[p(y_{ij}=1|\mathbf{u}_i)\right]^{y_{ij}}\left[1-p(y_{ij}=1|\mathbf{u}_i)\right]^{1-y_{ij}}\right)}\right)}\\
    &= \exp{\left(\sum_{j=1}^{n_i} y_{ij}\eta_{ij}\right)} \exp{\left(-\sum_{j=1}^{n_i}\log{(1 + \exp{(\eta_{ij})})}\right)},
\end{align}
where $\eta_{ij} = \mathbf{x}_{ij}^T\boldsymbol{\beta} + \mathbf{z}_{ij}^T\mathbf{u}_i$. Using the same reasoning as in Section \ref{polsuf}, $g(\mathbf{y}_i|\mathbf{u}_i;\boldsymbol{\beta})$ can also be expressed in terms of polynomial-approximate sufficient statistics per cluster $i$. As a result, the marginal log-likelihood function of the $i$th cluster
\begin{align}
    L_i(\boldsymbol{\beta}, \boldsymbol{\theta}) &= \int{g(\mathbf{y}_i|\mathbf{u}_i;\boldsymbol{\beta})f(\mathbf{u}_i;\boldsymbol{\theta})d\mathbf{u}_i},
\end{align}
and the full marginal likelihood
\begin{align}
    L &= \prod_{i=1}^N L_i(\boldsymbol{\beta,\theta}),
\end{align}
can be constructed from these polynomial-approximate sufficient statistics. Here, $f(\mathbf{u}_i;\boldsymbol{\theta})$ is the probability distribution of $\mathbf{u}_i$ with parameter vector $\boldsymbol{\theta}$. \\

Similar to the setting for a binary logistic regression model, our strategy for estimating the parameters of a mixed effects logistic regression entails generating the corresponding pseudo-data for the actual unavailable data through their summary statistics. This procedure is done for every cluster or data provider. 

\subsection{Motivating data}
To demonstrate the proposed strategy, we use the anonymized, time-shifted, and permuted publicly available data from the Children's Hospital of Pennsylvania (CHOP), which can be accessed through the \texttt{R} package \texttt{medicaldata} \citep{Rmedicaldata} \href{https://higgi13425.github.io/medicaldata/}{(https://higgi13425.github.io/medicaldata/)}. It consists of 15 524 COVID-19 test results along with patient information from 88 clinics of the hospital from days 4 to 107 since the start of the pandemic in 2020. The tests were performed via PCR and we are interested in modelling the association of the binary test result (negative or positive) and patient information. We include only complete cases and exclude clinics with only one patient record. We consider as predictors gender, patient class, whether or not the specimen was collected via drive-thru, age, and the number of days since the pandemic started. Gender and drive-thru indicator are binary covariates while patient class has more than two levels. We restrict the analysis to patients with patient class given as inpatient, emergency patient, and outpatient, because there are only a limited number of patients in the other levels. Note that outpatient comprises patients who were originally classified as \textit{recurring outpatient} and just \textit{outpatient}. Age and the number of days since the start of the pandemic are numeric variables which we standardized to minimize numerical difficulties during the estimation of the parameters and during the generation of pseudo-data, both of which involve nonlinear optimization \citep{JSSv067i01}. After pre-processing, the dataset used in this study comprised of 57 clinics with a total of 6 330 patient records.

\subsection{Simulation study}\label{sim}
 To investigate how much the estimates vary from the true parameter values based on the order of the moments matched, we compared the performance of estimates from pseudo-data that match up to the second, third, and fourth moments of the actual data through simulations. Matching up to only the second moment of the actual data requires the least information about the sensitive data and is thus the most preferred in terms of preserving privacy. However, matching up to the fourth moment has the potential advantage of replicating the log-likelihood of the actual data more closely (see a demonstration for varying coefficient values: \href{https://lizlimpoco.shinyapps.io/curvature_loglik/}{https://lizlimpoco.shinyapps.io/curvature\_loglik/}). Estimates from the actual data and the true parameter values served as bases for assessing the estimates in terms of bias and coverage. We used the COVID-19 test results data from CHOP as the basis for setting the parameters of the model.  \\

Using the same notations as in Section \ref{mixedglm}, we simulated data from a logistic regression model with random intercept per cluster such that:

\begin{align}
    Y_{ij} &\sim Bern(\pi_{ij}) \\
    X_1 &\sim N(\mu, \sigma^2) \\
    X_2 &\sim Pois(\lambda) \\
    \mathbf{X}_3 &\sim M_3(N; p_1, p_2, p_3) \\
    u_i &\sim N(0, \sigma^2_u)
\end{align}

\begin{align}
    \text{logit} (\pi_{ij}) &= \beta_0 + \beta_1x_{1_{ij}} + \beta_2x_{2_{ij}} + \beta_{32}x_{{32}_{ij}} + \beta_{33}x_{{33}_{ij}} + u_i \\
    \sigma_u = 1.195, \beta_0 &= -4.16, \beta_1 = 0.32, \\
    \beta_2 = -0.24, \beta_{32} &= 1.20, \beta_{33} = 0.74,
\end{align}

where we denote the total number of observations across all clusters as $N$. Aside from the two numeric variables whose distributional parameters are based on the CHOP variables namely age and number of days since the start of the pandemic, we included a categorical variable $\mathbf{X}_3$ with three levels. This latter variable has distributional parameters based on the variable patient class. Three combinations of number of clusters ($m$) and cluster sizes ($n_i$) were considered, namely: (1) $m = 30$ and $n_i = 100$; (2) $m = 50$ and $n_i = 60$; and (3) $m = 100$ and $n_i = 30$. In all cases, $N = 3000$ and the number of observations per cluster is the same ($n_i = n$). \cite{Moineddin2007} recommended at least 50 clusters with at least 50 observations per cluster to produce reliable estimates. Setting 2 follows this recommendation while Settings 1 and 3 are chosen to explore the effect of a small number of clusters or cluster size, respectively, on the estimates produced from pseudo-data.  \\

For each case, 100 sets of data were simulated for both the predictors and response. These simulated data act as the ``actual data" from which summary statistics were computed. Corresponding pseudo-data matching up to the second, third, and fourth moments were then generated based on the summary statistics. Models based on the simulated and pseudo-data were estimated using the \texttt{glmer} function from the \texttt{R} package \texttt{lme4} \citep{Rlme4}. \cite{Austin+2010} conducted simulations and presented that among the likelihood-based procedures for estimating multilevel logistic regression, the Gauss-Hermite quadrature implemented by \texttt{glmer} provides superior performance whenever the number of clusters is small (5 to 20) and had no substantial difference in performance compared to the other methods for large samples.

\section{Simulation results}
Non-convergence messages (warnings or errors) were generated for 39\%, 17\%, and 0\% of the 100 simulated datasets in each setting, respectively, when computing point estimates of the parameters. On the other hand, the non-convergence rates were 42\%, 28\%, and 7\% in each setting, respectively, when estimating 95\% confidence intervals via profile likelihood. According to the documentation of the \texttt{R} package \texttt{lme4} \citep{Rlme4}, convergence warnings may be false positives which can be attributed to the difficulty of assessing convergence of the optimization algorithms implemented. The authors of the package considered it a gold standard to run the estimation procedure using all available optimizers and then to check whether the estimates are practically equivalent. However, even though the point estimates produced by the various optimizers differed only in the first digit after the decimal point at most, thereby indicating a false non-convergence warning, the interval estimates computed via profile likelihood still generated errors. Particularly, these errors pertain to the detection of deviances lower than that achieved at the supposedly optimum parameter estimates. \cite{JSSv067i01} explained that this error can occur if the initial fit is wrong due to numerical problems. 
\cite{Moineddin2007} also encountered non-convergence in their simulation experiments, which significantly improved with an increase in the number of groups or group size. They noted that low prevalence outcomes require larger group sizes and briefly explained that when the sample size is small there may not be sufficient variation to estimate random effects, running the risk of non-convergence. Indeed, when we explored additional settings with the same number of clusters but with an increased cluster size of 200, non-convergence was no longer an issue either for the simulated or pseudo-data (Appendix \ref{app:additional_settings}).  \\

\begin{figure}[h]
    \centering
    \includegraphics[width=1\linewidth]{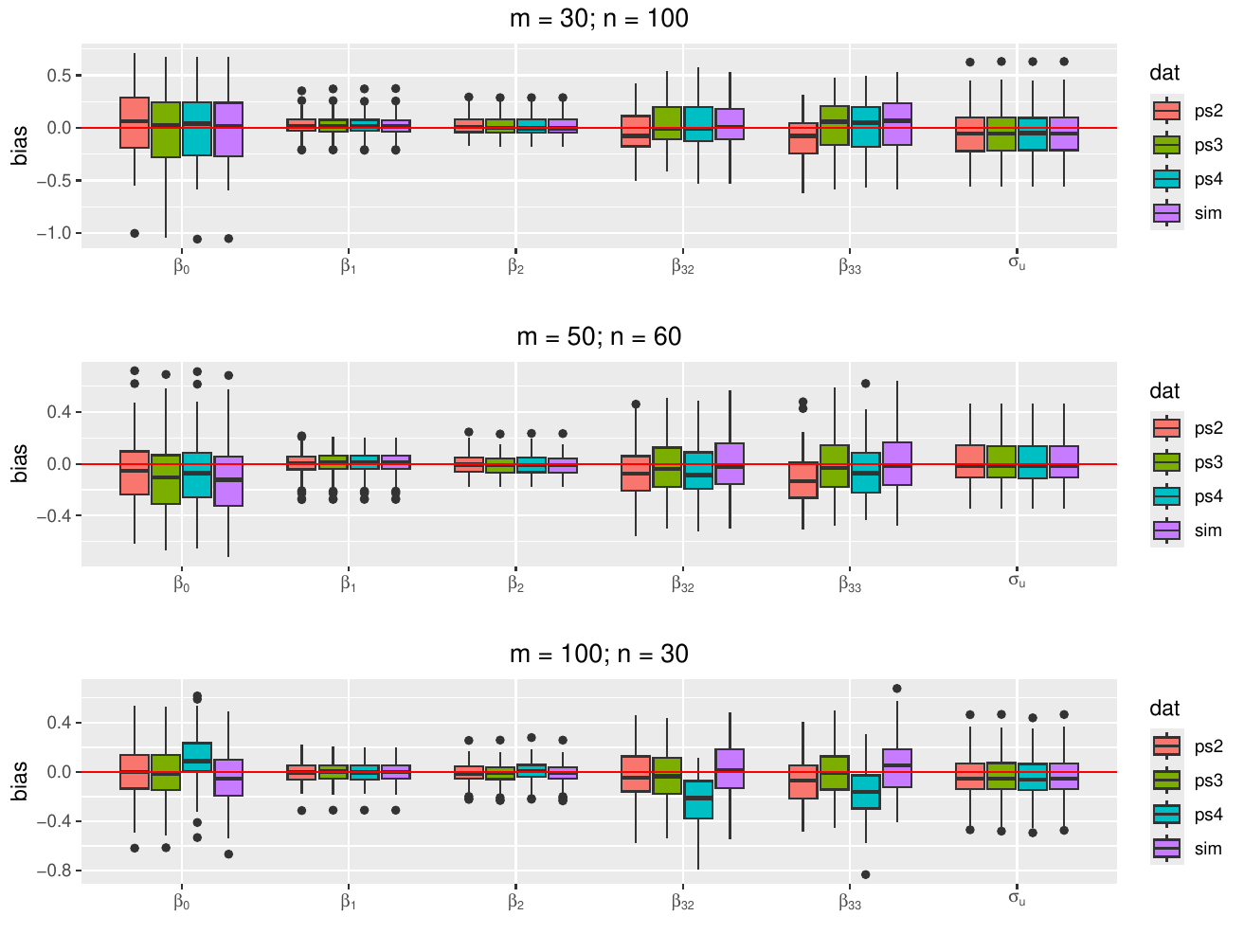}
    \caption{Comparing the bias distribution of estimates derived from pseudo-data matching up to the second moment only (ps2), up to third (ps2), and fourth (ps4) moments with those derived from the simulated data (sim) across 100 sets of data for different combinations of number of clusters $m$ and cluster size $n$.}
    \label{fig:bias}
\end{figure}

We compared the bias distribution of the estimates derived from the simulated and pseudo-data across 100 different sets for the six parameters: the standard deviation of the random intercepts $\sigma_u$ and the five fixed effects $\beta_0$, $\beta_1$, $\beta_2$, $\beta_3$, and $\beta_4$ (Figure \ref{fig:bias}). We observe that the bias distributions for the numerical covariate effects $\beta_1$ and $\beta_2$ as well as for $\sigma_u$ are very similar across all settings, whether the actual dataset or pseudo-dataset was used. Estimates derived from pseudo-data matching up to the second moment only performed relatively poorly for covariate effects with larger [absolute] true values such as $\beta_{32} = 1.20$ and $\beta_{33} = 0.74$ especially when the cluster size is relatively large ($n=60$ and $n=100$). On the other hand, the bias distribution of estimates derived from pseudo-data matching up to the fourth moment deviates more from the others whenever the cluster size is small ($n=30$ and $n=60$) and the true parameters are relatively large ($\beta_{32}$ and $\beta_{33}$). Pseudo-data that matches up to the third moment of the simulated data appear to produce estimates that perform as good as those based on the simulated data in all settings and parameters in terms of the bias distribution. \\  

\begin{figure}[h]
    \centering
    \includegraphics[width=1\linewidth]{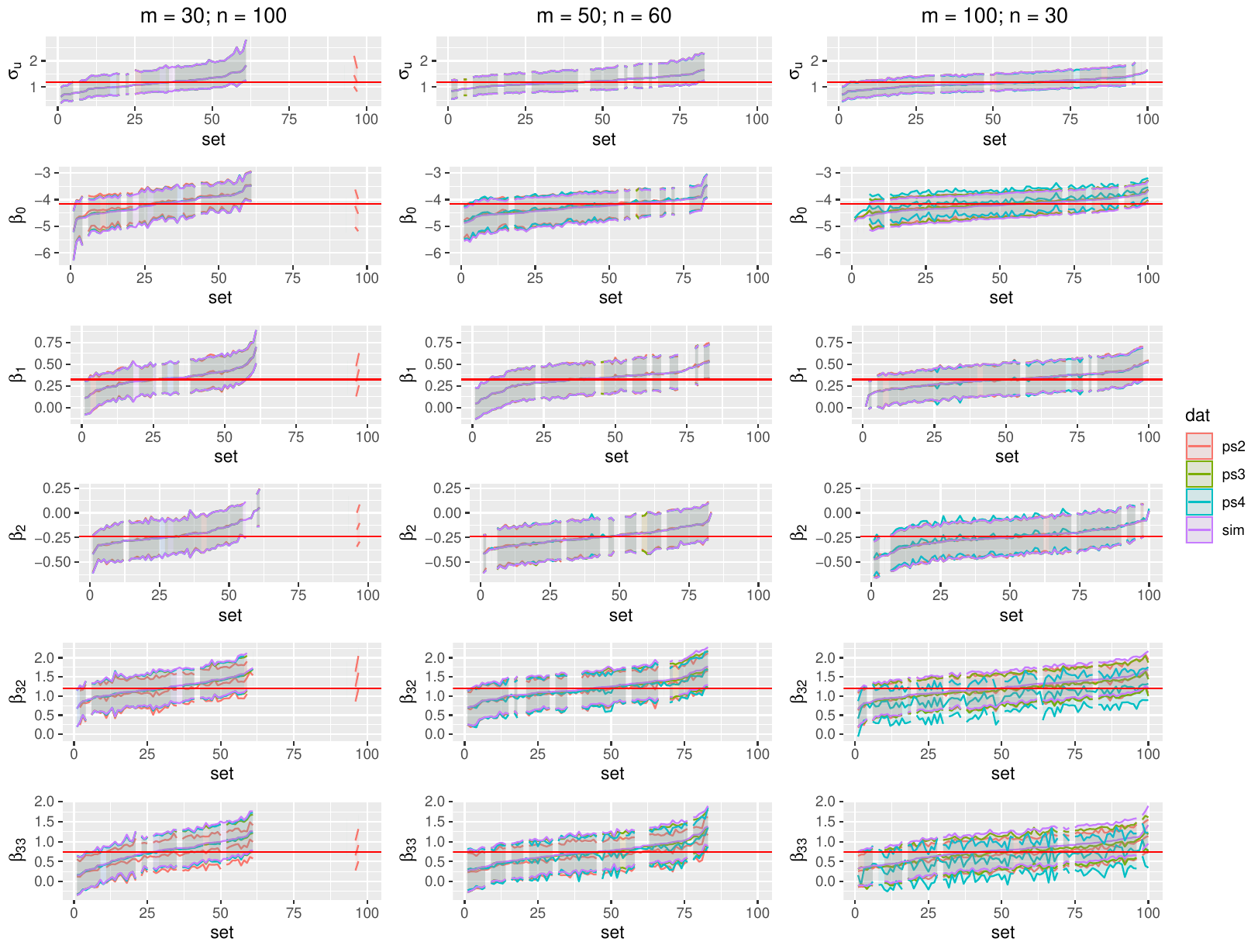}
    \caption{95\% Confidence interval estimates (upper and lower curves) of all parameters (y-axis) derived from the pseudo-data matching up to the second, third, and fourth moments (ps2, ps3, and ps4, respectively) compared with those derived from simulated data (sim) across 100 datasets (x-axis). The red horizontal line indicates the true parameter value while the middle curve represents the corresponding point estimates. Due to convergence issues in some sets, estimates are not available for the simulated or pseudo-data.}
    \label{fig:confint}
\end{figure}

A similar pattern is observable with regards to the 95\% confidence interval estimates produced from the simulated and pseudo-data (Figure \ref{fig:confint}). Pseudo-data matching up to the second moment only (ps2) produces interval estimates that are visibly different from those derived from the simulated data (sim) when estimating relatively large covariate effects ($\beta_{32}$ and $\beta_{33}$). Likewise, pseudo-data matching up to the fourth moment (ps4) also has difficulty mimicking the interval estimates derived from simulated data for larger covariate effects but only when the cluster size is small ($n=30$). \\

\begin{figure}[h]
    \centering
    \includegraphics[width=1\linewidth]{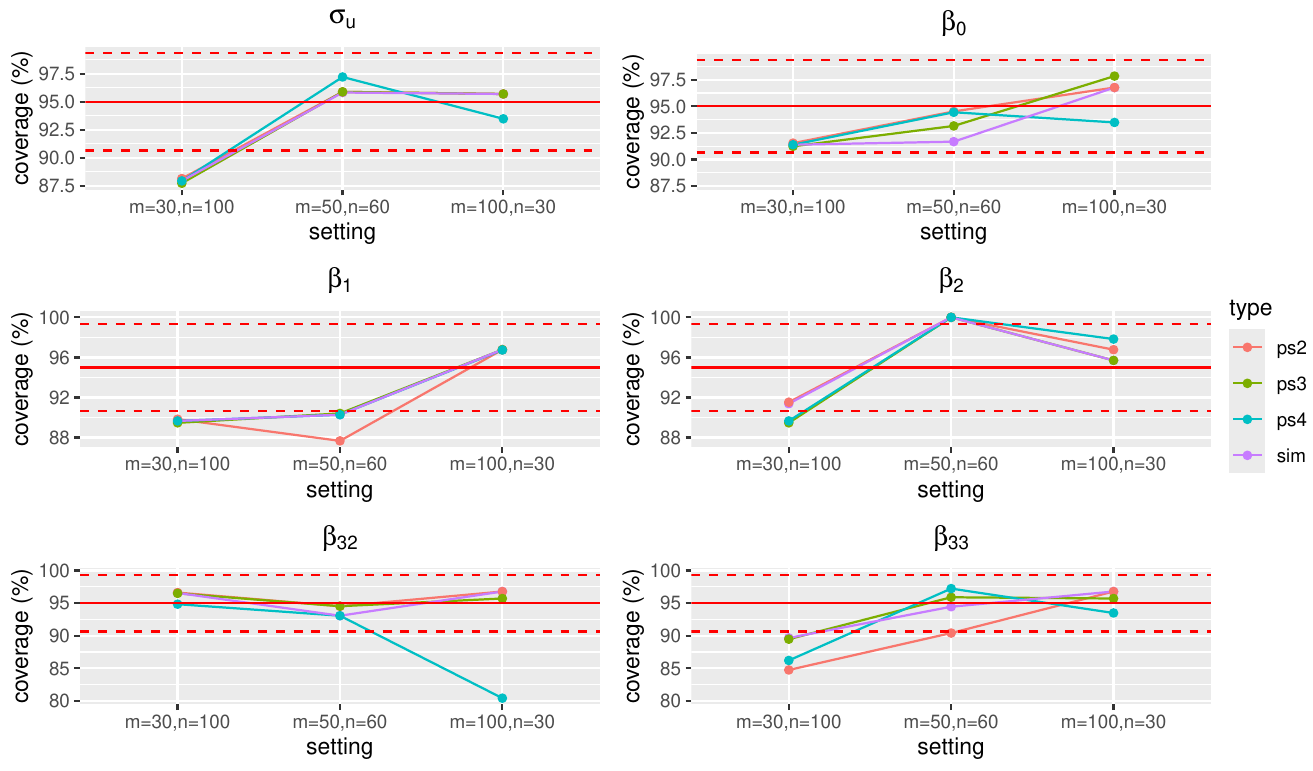}
    \caption{95\% confidence interval coverage of estimates derived from pseudo-data matching up to the second, third, and fourth moments (ps2, ps3, and ps4, respectively) compared with those derived from the simulated data (sim). The red solid horizonal line indicates the nominal coverage ($95\%$) while the red dashed horizonal lines indicate the lower and upper values within which the coverage is considered acceptable \citep{https://doi.org/10.1002/sim.2673}}
    \label{fig:coverage}
\end{figure}

Figure \ref{fig:coverage} displays the 95\% confidence interval coverage across the different settings using the simulated data and pseudo-data. Here we see the impact of interval estimate deviations due to utilizing the pseudo-data instead of the simulated data. \cite{https://doi.org/10.1002/sim.2673} suggested to assess acceptability of coverage based on whether it falls within approximately 2 standard errors of the nominal coverage probability ($p$); that is
\begin{align}
    SE(p) &= \sqrt{p(1-p)/B},
\end{align}
where $B$ is the number of simulations. In our study, $B=100$ and $p=95\%$. Hence, for the coverage to be considered acceptable, it must fall between $[90.64\%,99.36\%]$. From the plots, notable under-coverage occurs when using pseudo-data matching up to the second moment for the parameter $\beta_{33}$ and up to the fourth moment for $\beta_{32}$. For $\sigma_u$, under-coverage only occurred in the first setting ($m=30,n=100$) wherein even the coverage arising from the simulated data is considered unacceptable. No under- nor over-coverage is observed for $\beta_0$, although there are noticeable differences in coverage in the second and third settings. On the contrary, coverage for $\beta_1$ is almost identical in all settings for estimates derived from the simulated data and pseudo-data, except for estimates from pseudo-data matching up to the second moment only. Finally, $\beta_2$ coverage does not vary much when using simulated data nor pseudo-data across different settings. \\

Model selection is an integral part in any statistical modelling process. The Akaike Information Criterion (AIC) is one of the most popular measures used to select a model from a set of candidates \citep{doi:10.1080/1743727X.2014.986027}. It is computed based on the maximum log-likelihood ($\log{(L(\hat{\boldsymbol{\theta}}|\mathbf{y}))}$) and the number of parameters in the model $K$ \citep{burnham2003model}
\begin{align}
    AIC &= -2\log{(L(\hat{\boldsymbol{\theta}}|\mathbf{y}))} + 2K.
\end{align}
We examined the AIC values computed from the models estimated based on the pseudo-data and compared them with those derived from the actual data (Appendix \ref{app:aic}). We observe very small differences, if any, except in the setting where there are $m=100$ clusters each with a sample of $n=30$. In this setting, the AIC values based on pseudo-data matching up to the fourth moment are quite distinguishable from those based on the actual data.  \\

Overall, the simulation results suggest that pseudo-data matching up to the third moment of the simulated data achieve the most similar performance with the simulated data in terms of bias, coverage, and AIC values across all settings and parameters.

\section{Illustrative example: COVID-19 data from CHOP}
Suppose there is interest in determining how patient characteristics affect COVID-19 incidence. Raw patient data have been collected from all clinics in CHOP but for privacy purposes, these cannot be fully disclosed to another party without the necessary process and paperwork, which may take considerable time and effort. However, the data custodian of CHOP may be willing to share a summarized version of the data more easily than the individual-level records. In this section we illustrate how the proposed strategy applies to a real-world setting with two parties: the data provider and the data analyst. We outline the tasks and information expected from both when applying the approach. Relevant \texttt{R} codes are available in our Github repository \href{https://github.com/lizlimpocouhasselt/Mixed-effects-logistic-regression-from-summary-statistics/}{(https://github.com/lizlimpocouhasselt/Mixed-effects-logistic-regression-from-summary-statistics/)}.  

\subsection{Required summary statistics from data providers}
The proposed approach expects summary statistics from each of the 57 clinics at CHOP. Based on the results of the simulation experiments, summary statistics comprised of the sample moments up to the third order (including joint multivariate moments) of the actual data would be polynomial-approximate sufficient. For demonstration purposes though, we consider generating pseudo-data matching up to the fourth sample moments and so we compute the sample moments up to the fourth order for each clinic. 

\subsection{Data analyst's tasks}
From the summary statistics per clinic, the data analyst employs the proposed two-step procedure: pseudo-data generation and model estimation. 
\subsubsection{Pseudo-data generation}
\begin{figure}[h]
    \centering
    \includegraphics[width=1\linewidth]{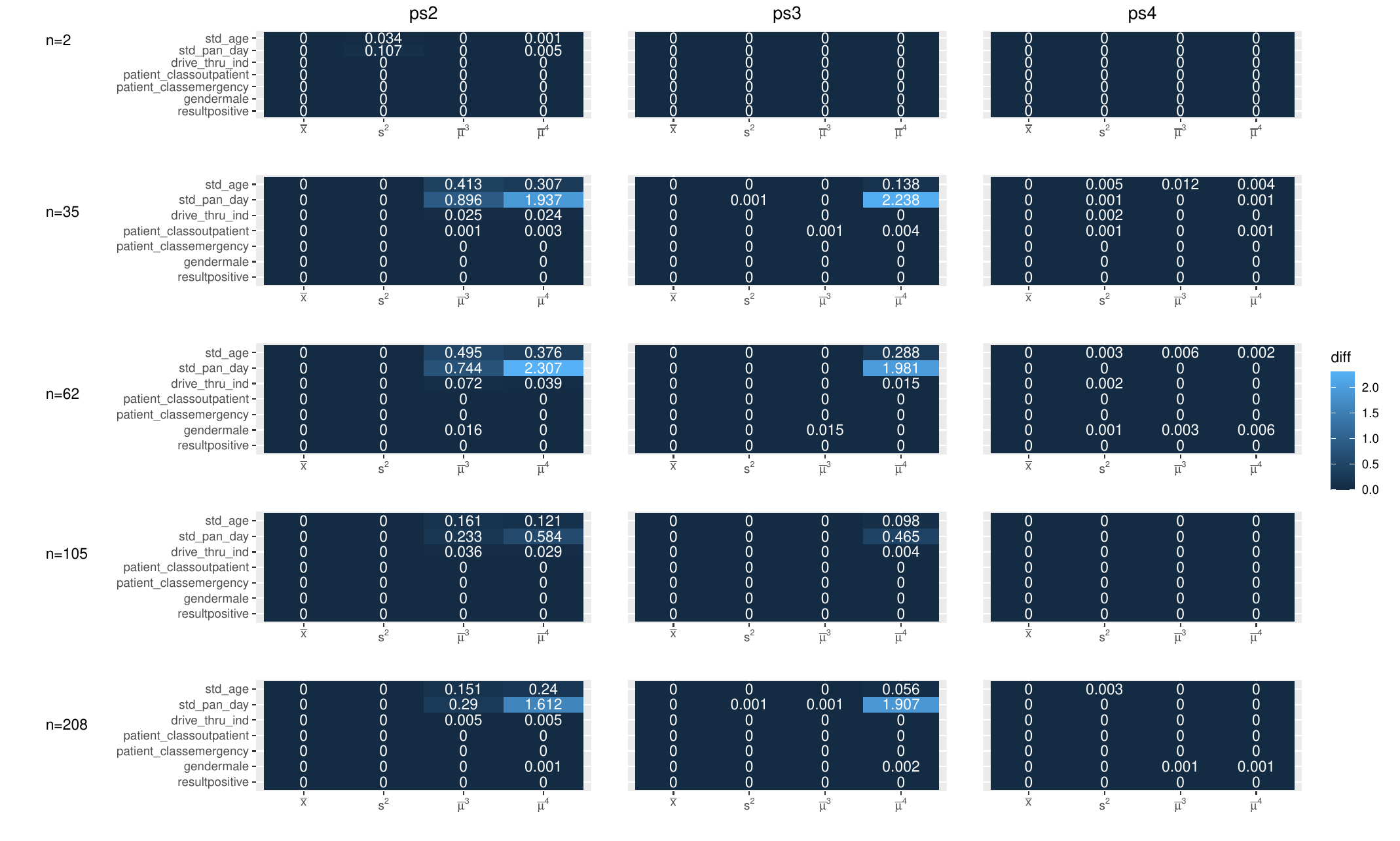}
    \caption{Absolute difference in univariate sample moments between pseudo-data matching up to the second (ps2), third (ps3), and fourth (ps4) moments and the actual CHOP data.}
    \label{fig:sumstat_ps}
\end{figure}

Generation of pseudo-data was implemented as described in Section \ref{genpseudo}. Figure \ref{fig:sumstat_ps} displays the absolute difference in univariate sample moments between the generated pseudo-data based on different orders (up to second, third, or fourth) and the actual data from 5 clinics at CHOP namely \textit{3 Laboratory} ($n=2$), \textit{Inpatient Ward D} ($n=35$), \textit{Inpatient Ward E} ($n=62$), \textit{Inpatient Ward H} ($n=105$), and \textit{Inpatient Ward A} ($n=208$). These clinics were selected among the 57 clinics at CHOP to illustrate how cluster sample size affects the matching of sample moments between the pseudo- and actual data. For instance, for clinics with small samples such as \textit{3 Laboratory} ($n=2$), \textit{Inpatient Ward D} ($n=35$), and \textit{Inpatient Ward E} ($n=62$), matching up to the fourth sample moments (3rd column) of the actual data is more difficult, whereas matching only up to the second or third order is manageable. However, with large enough samples such as for \textit{Inpatient Ward H} ($n=105$) and \textit{Inpatient Ward A} ($n=208$), this is no longer observed.

\subsubsection{Model estimation}
Once the pseudo-data have been generated, model estimation proceeds as if the pseudo-data are the actual data. Based on the simulation results, since estimates from pseudo-data matching up to the third sample moments of the actual data perform as good as those derived from actual data amid varying cluster sizes and true parameter values, we use this artificially generated dataset to estimate a mixed effects binary logistic regression for the CHOP data. Table \ref{tab:chop_glmm_ps3_vs_actual} presents the comparison of the models fitted on the pseudo-data and the actual data where we see that the estimates are similar except for the scaled residuals, which we cannot replicate unless we have access to the actual observations.

\begin{table}[h]
\caption{Comparison of mixed effects binary logistic regression models with random intercept only based on pseudo-data and based on actual CHOP data.}
%\begin{left}
\resizebox{\textwidth}{!}{\color{black}
\begin{tabular}{lrrrr}
\hline
 & \multicolumn{2}{c}{pseudo-data (3rd)} & \multicolumn{2}{c}{actual data} \\
 & Est(std. err.) & 95\% CI & Est(std. err.) & 95\% CI \\
\hline
(Intercept) & -4.169 (0.322){***} & (-4.8917, -3.5735) & -4.173 (0.327){***} & (-4.9143, -3.5943) \\
Gendermale & -0.167 (0.123){\textcolor{white}{***}} & (-0.4079, 0.0729 ) & -0.163 (0.123){\textcolor{white}{***}} & (-0.4046, 0.0768 )\\
Patient class (emergency) & 1.197 (0.181){***} & ( 0.8499, 1.5593 ) &1.214 (0.182){***}& ( 0.8650, 1.5808 )\\
Patient class (outpatient) & 0.505 (0.376){\textcolor{white}{***}} & (-0.2284, 1.2732 )& 0.536 (0.374){\textcolor{white}{***}}&(-0.1940, 1.3015 ) \\
Drive thru ind & 0.250 (0.253){\textcolor{white}{***}} & (-0.2711, 0.7208 ) & 0.328 (0.245){\textcolor{white}{***}} & (-0.1776, 0.7860 )\\
Std. Pandemic Day & -0.252 (0.065){***} & (-0.3786, -0.1248) & -0.253 (0.064){***} & (-0.3800, -0.1276)\\
Std. Age & 0.363 (0.052){***} & ( 0.2625, 0.4651 ) & 0.338 (0.045){***} & (0.2461, 0.4350 )\\
$\sigma_{Int}$ & 1.078{\textcolor{white}{***}}&  ( 0.7340, 1.6231 ) & 1.076{\textcolor{white}{***}}     &( 0.7312, 1.6218 )     \\
\hline
 Scaled residuals: & & \\
 \hspace{1cm}{Min} & -5.296 & & -1.627\\
 \hspace{1cm}{$Q_1$} & -0.278 && -0.289\\
 \hspace{1cm}{Median} & -0.161 && -0.164\\
 \hspace{1cm}{$Q_3$} & -0.113 && -0.112\\
 \hspace{1cm}{Max} & 13.898  && 10.538\\
AIC                           & 2211.3    &          & 2210.4              \\
BIC                           & 2265.3     &         & 2264.4              \\
N                     & 6330           &         & 6330                    \\
number of clinics     & 57             &          & 57                       \\
\hline
\multicolumn{3}{l}{\footnotesize{$^{***}p<0.001$; $^{**}p<0.01$; $^{*}p<0.05$}}
\end{tabular}}

\label{tab:chop_glmm_ps3_vs_actual}
%\end{left}
\end{table}

\section{Discussion}
The findings from the simulation experiments indicate that the pseudo-dataset matching up to the third sample moments of the actual data is a viable substitute to the actual data. Although pseudo-data matching up to the second moment only is more convenient, estimates produced from these are more easily prone to bias and under-coverage if the true parameter values are relatively large. This can be attributed to the error term when re-expressing the log-likelihood using Taylor's theorem (Section \ref{polsuf}). This error term defines the difference between the log-likelihood based on the actual data and based on pseudo-data such that increasing $\boldsymbol{\beta}$ values result in more divergence between the log-likelihoods. Consequently, if the true parameter value is relatively large, the optimum $\boldsymbol{\beta}$ will also be relatively large, at which point the gap between the log-likelihoods becomes large enough to return different optima. Although the pseudo-data matching up to the third and fourth sample moments of the actual data also suffer from the same limitation, pseudo-data matching up to the second moment suffer the most even with small log-odds covariate effects. Further simulation experiments (not presented here) showed that estimates from pseudo-data matching up to the third and fourth sample moments are more robust whenever the true [absolute] log-odds effects are as high as 2.5. In practice, knowledge about the magnitude of covariate effects is often unavailable. Thus, a safer choice would be to use pseudo-data that matches at least up to the third moment. \\

A relevant concern when using pseudo-data that matches at least up to the third moment pertains to cases when the true parameter values are extremely large, i.e. more than 2.5. It is important to note that the parameter values, although they can be any real number, they are limited to their practical interpretation in real-world settings, especially in the case of a logistic regression model. For instance, large absolute values for the intercept result from very rare or very common events, such as the case for the CHOP data whose estimated intercept is $-4.17$, which reflects the low prevalence of positive COVID-19 test outcomes in the dataset ($4.8\%$). True absolute intercept values larger than this do not occur often as this already refers to much rarer events that even the estimates produced from actual observations are biased \citep{doi:10.1080/02664763.2017.1282441,doi:10.1177/10944281221083197}. Moreover, the estimates produced through our proposed method are not as heavily affected by the size of the intercept as they are by the size of the covariate effects. This can be deduced by examining the same error term mentioned previously. With regards to the covariate effects, a true value of 3 is equivalent to an odds ratio of about 20, which is already quite huge for a true variable effect size, and might not always exist from real-world data especially with multiple variables in the model. \\

Our proposed pseudo-data generation is implemented through solving a nonlinear least squares optimization problem, which offers some advantages. One benefit is that we do not sample from a distribution; instead, the pseudo-data are produced by searching the solution from an initial point along the objective function until convergence. As a consequence, no distributional assumption is necessary. Another advantage is the improbability of generating the exact actual data, thereby upholding the privacy of the actual sensitive data. This is explained by the infinitely many solutions to a nonlinear least squares problem whenever there are more pseudo-data points to be generated than there are sample moments to match. For example, to generate pseudo-data that matches up to the third sample moments (univariate and multivariate) for the CHOP data with seven variables composed of a response variable and six predictors, summary information should comprise: a $7 \times 1$ mean vector; a $7 \times 7$ symmetric covariance matrix; two $6 \times 6$ symmetric matrices of univariate and bivariate joint sample central moments of order 3 (excluding the response variable); and a $\binom{6}{3} = 20 \times 1$ vector of trivariate joint sample central moments of order 3. In total, 97 summary measures must be matched between the pseudo-data and actual data. If the actual data for a clinic has more than 97 observations, say $n=105$ like in the case of \textit{Inpatient Ward H}, it follows that the size of the solution to the optimization problem is also $n=105$, which equates to the size of the pseudo-data to be generated. In this case, the problem is underdetermined which results in infinitely many solutions. Under this setting, it is easier for the optimization algorithm to find a solution that almost, if not exactly, matches the specified summary statistics. \\

In the overdetermined case wherein the size of the solution is smaller than the available information, the algorithm may have difficulty satisfying all the summary measures that need to be matched, reducing the similarity between the summary statistics of the pseudo-data and the actual data. This explains why matching the pseudo-data up to the fourth moments of the actual data appears to be less effective whenever the cluster size is small. Such is the case for \textit{Inpatient Ward D} whose cluster size is just $n=35$. Consequently, the performance of the estimates are affected, as was observed in the simulation experiments for the setting where the cluster size is $n=30$. 
\\

Because the pseudo-data have statistical properties similar to the actual data with respect to the models of interest, once they are generated, the modelling process proceeds like with the actual data. In contrast, directly using the summary statistics requires re-expressing the log-likelihood function as well as the other relevant estimators such that their input should require aggregated information instead of the classical individual-level data. When extending to other statistical models, this is not trivial especially when random effects are involved. This is an edge of our proposed strategy, which facilitates modelling and the possibility of generalising to other statistical models such as those in the GLM family. Along with the work of \cite{limpoco2024linearmixedmodellingfederated} which uses the same strategy for linear mixed models, our approach for logistic mixed models provides a framework for dealing with model estimation under privacy restrictions. In practice, the statistical modelling workflow may then consist of pseudo-data generation and then estimation of either a linear or logistic mixed model, depending on the model of interest. If the data analyst has decided to estimate a linear mixed model beforehand, pseudo-data generation that matches up to the second moment only of the actual data is simpler and faster. \\

Aside from generalisability, pseudo-data generation also facilitates model selection via AIC or other log-likelihood-based criteria. \cite{burnham2003model} emphasised against the pitfall of comparing AIC values of models derived from different datasets. They cited an example of comparing the AIC of a model obtained from a particular dataset with the AIC of a model estimated from the same dataset but with the outliers removed. This analysis, according to them, is invalid. In our case under the data privacy setting, we do not have access to the actual data to estimate models from them. Thus, our proposal involves generating pseudo-data as a substitute to the model estimation process and AIC values from models estimated on the same pseudo-dataset only can be compared. Our simulation results indicate that pseudo-data-based models closely resemble those that would have been derived, had the actual data been available. It follows that the model selection results will also be similar. We conducted simulation experiments on this (not shown here) and observed that the results support this hypothesis. \\

The nature of our proposed approach does not depend on the estimation method, unlike the work of \cite{10.1093/jamia/ocac067} which relies on the competence of the Penalized Quasi-Likelihood (PQL) method. It enables collaboration among more data providers since transfer of summary information is easier, faster, and does not require iterative communication nor infrastructure while maintaining privacy of the actual data. In addition, existing model estimation functionalities in any statistical software may be utilized once the pseudo-data are generated. Overall, the proposed strategy provides an excellent option to tackle the challenge of statistical modelling of sensitive data. \\

The proposed method makes use of the Levenberg-Marquardt algorithm to generate pseudo-data, although there might be other more efficient and effective optimization algorithms which can achieve the same purpose. We have found that generating pseudo-data one variable at a time significantly helps in minimizing the computational time, but it has the drawback of having difficulty matching the summary statistics between the pseudo-data and actual data for variables that come last. Hence, it would be interesting to find other approaches that do not suffer from this weakness. \\

\section{Conclusion}
In this study, we proposed an alternative approach to federated learning when modelling a mixed effects binary logistic regression. Instead of requiring more than one communication iterations and infrastructure, our proposed strategy only requires data providers to share polynomial-approximate sufficient statistics once without the need for secure database connections between the data providers and data analyst. We propose to generate pseudo-data from these polynomial-approximate sufficient statistics, which we found to be best comprised of the sample central moments up to the third order of the actual data. With the two-step procedure of pseudo-data generation and model estimation, we are able to estimate the model as reliably as one which utilizes the actual data.  

\section{Declaration of conflicting interests}
The author(s) declare(s) no potential conflicts of interest with respect to the research, authorship and/or publication of this article.

\section{Acknowledgements}
This study was supported by the Special Research Fund of Hasselt University (BOF24OWB22, Methusalem grant).

\appendix
\section{Appendix}
\subsection{Bias, coverage, and AIC for settings with $n=200$ observations per cluster}\label{app:additional_settings}
\begin{figure}[H]
    \centering
    \includegraphics[width=1\linewidth]{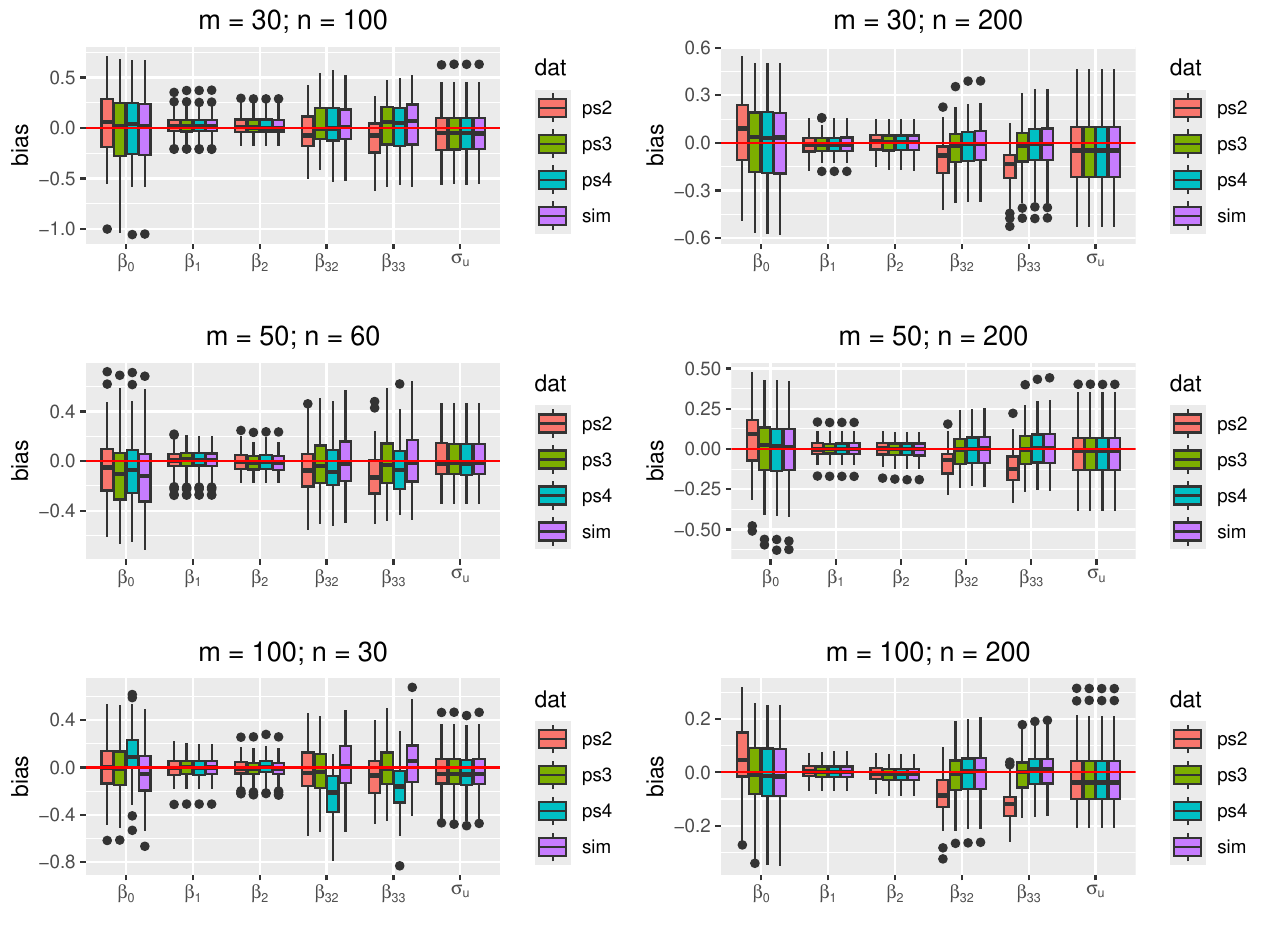}
    \caption{Bias distribution across 100 sets of data for different combinations of number of clusters $m$ and cluster size $n$.}
    \label{fig:bias6settings}
\end{figure}

\begin{figure}[H]
    \centering
    \includegraphics[width=1\linewidth]{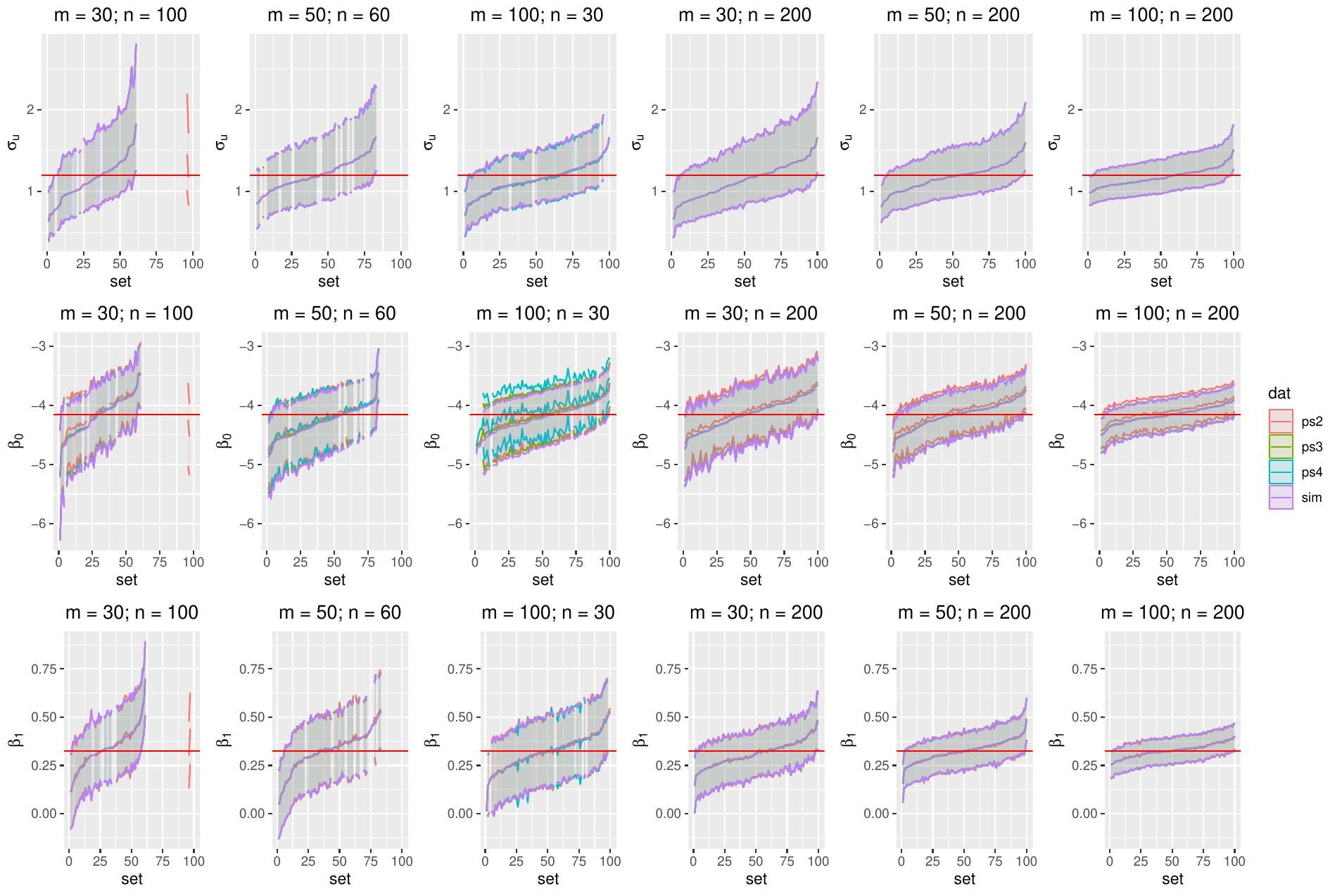}
    \includegraphics[width=1\linewidth]{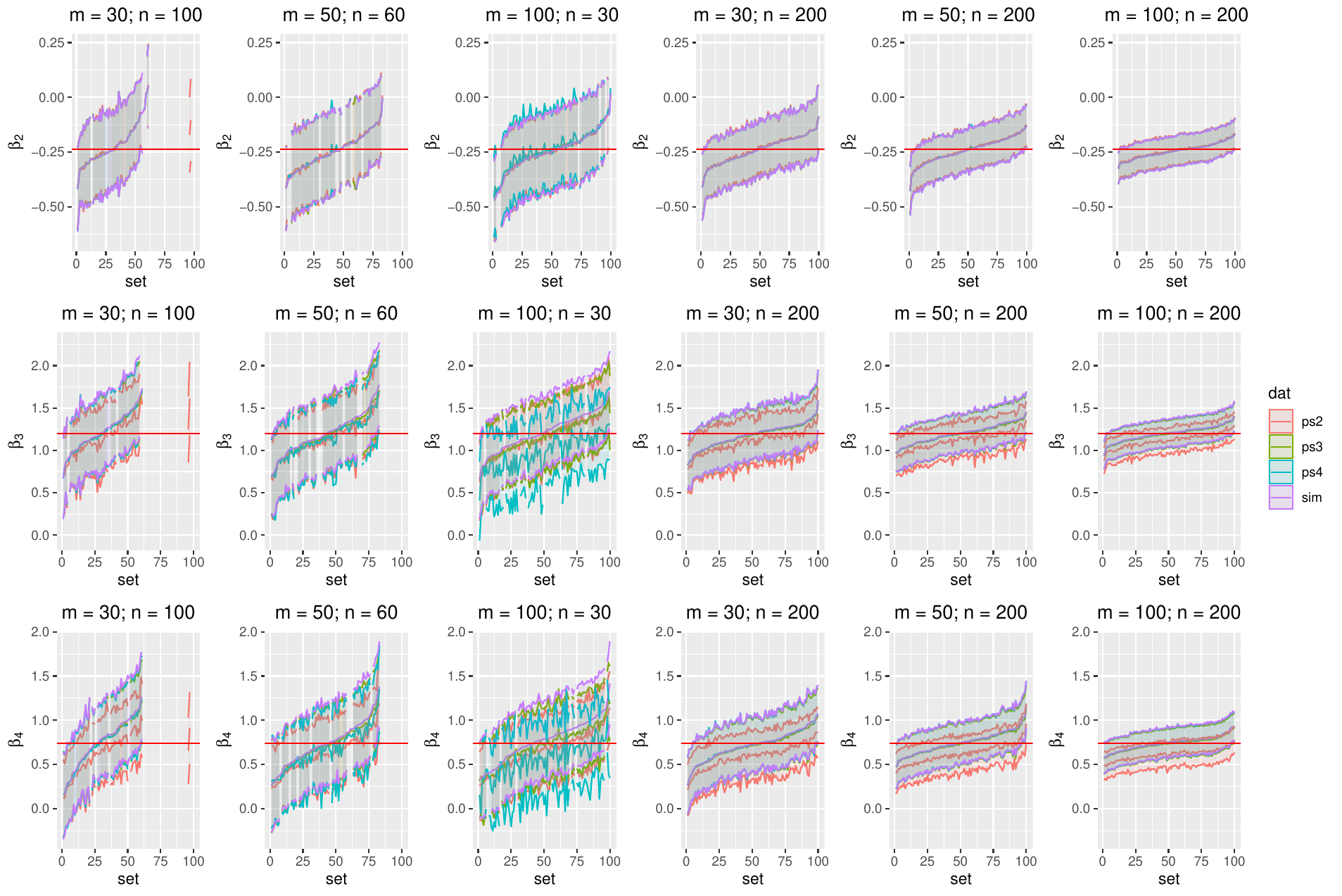}
    \caption{95\% Confidence intervals}
    \label{fig:confint6settings}
\end{figure}

\begin{figure}[H]
    \centering
    \includegraphics[width=1\linewidth]{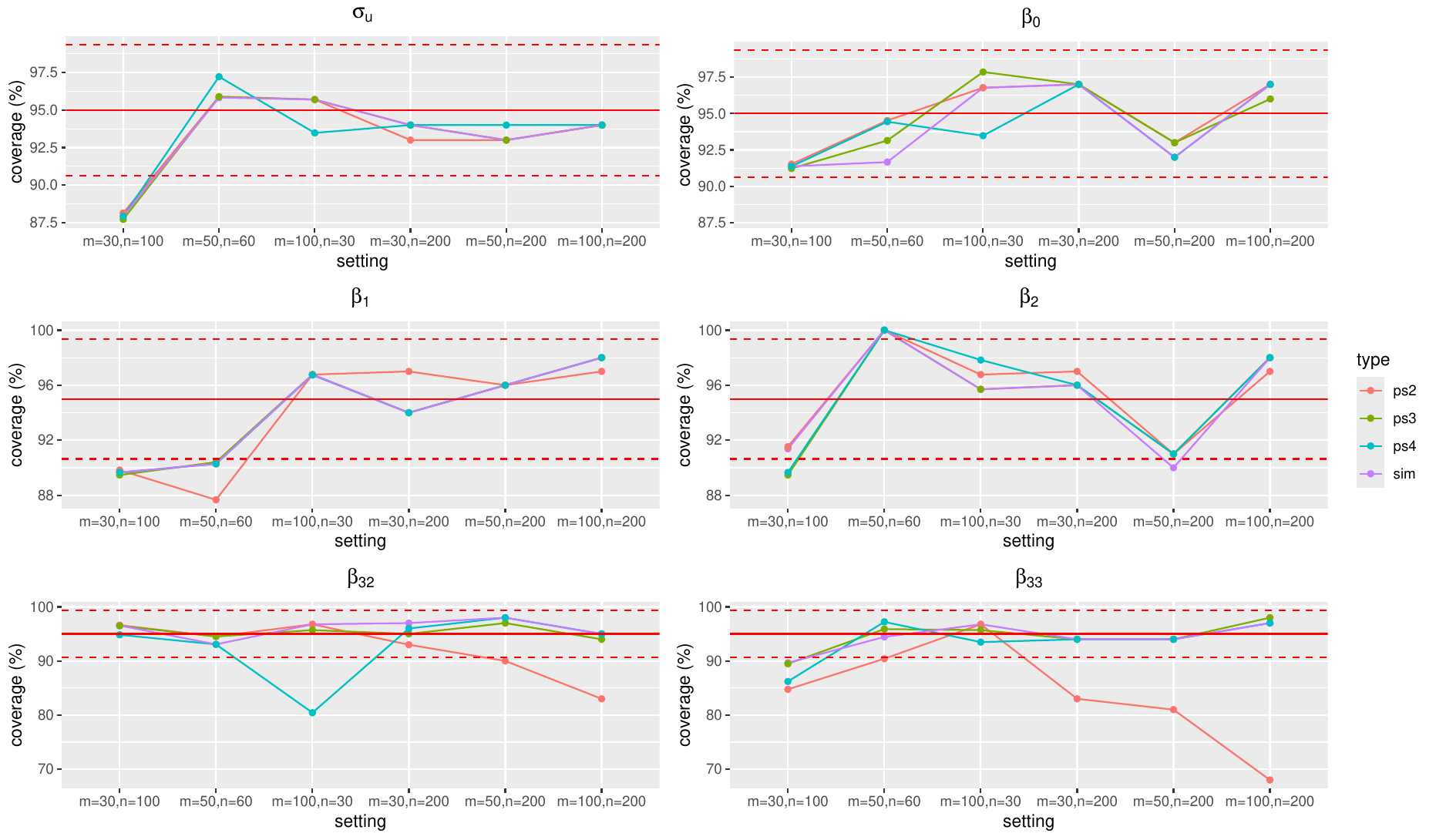}
    \caption{95\% confidence interval coverage}
    \label{fig:coverage6settings}
\end{figure}

\subsection{Comparison of AIC values computed from the pseudo- and actual data}\label{app:aic}
\begin{figure}[H]
    \centering
    \includegraphics[width=1\linewidth]{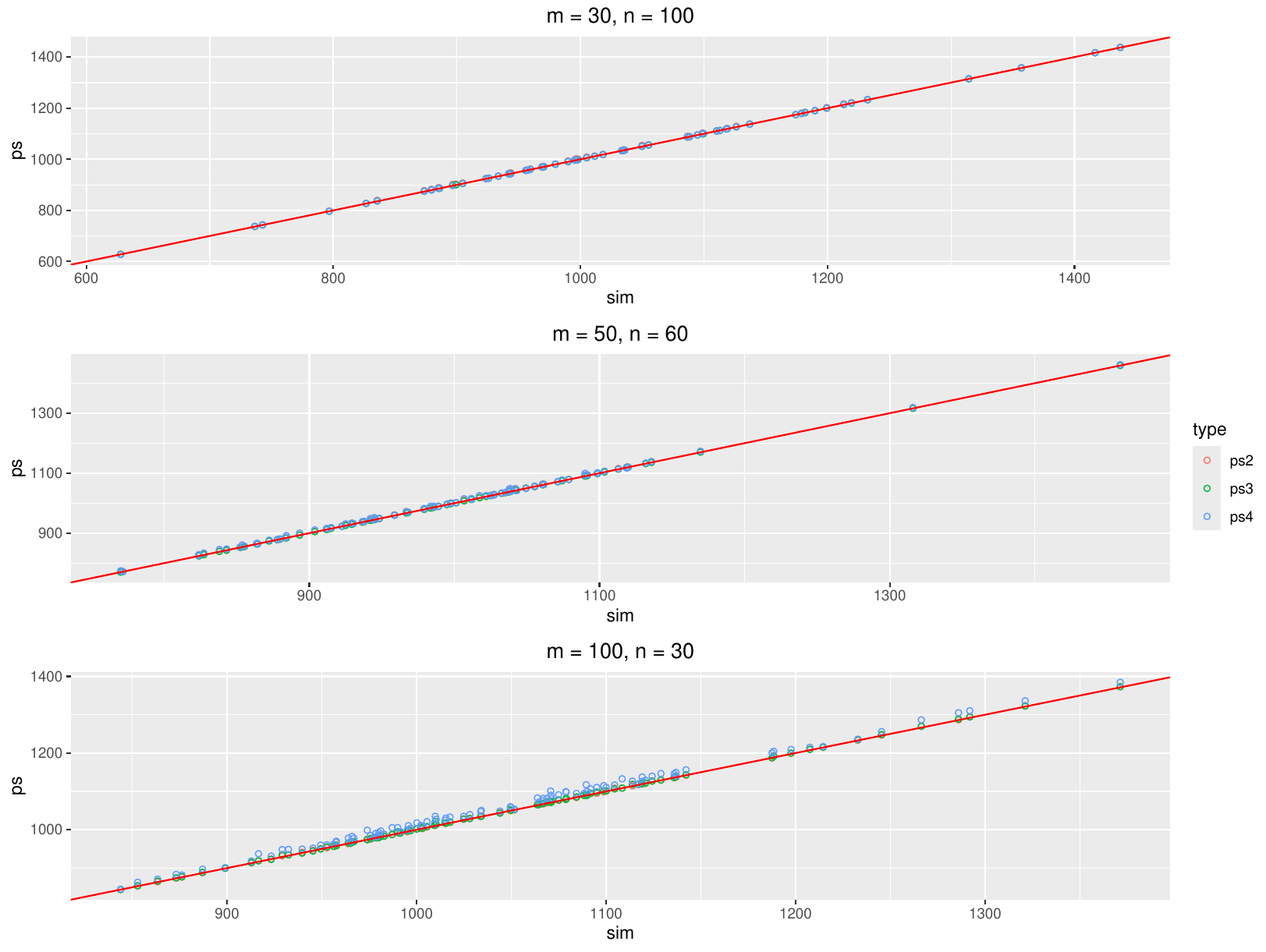}
    \caption{Comparison of AIC values between those derived from simulated data (x-axis) and those derived from pseudo-data matching up to the second (ps2), third (ps3), and fourth (ps4) moments. Points above the red diagonal line indicate AIC values from pseudo-data that are higher than those from the simulated data while those below the red line indicate lower AIC values compared to those derived from the simulated data.}
    \label{fig:aic}
\end{figure}

\bibliographystyle{abbrvnat}
\bibliography{Ref/ref}

\hypersetup{linkcolor=cyan}

\appendix

\end{document}